# Incorporating global dynamics to improve the accuracy of disease models: Example of a COVID-19 SIR model

Hadeel H. AlQadi[1,2], Majid Bani-Yaghoub[1]

[1] Department of Mathematics and Statistics, University of Missouri-Kansas City, Kansas City, Missouri, United States of America.

[2] Department of Mathematics, Jazan University, Saudi Arabia.

## Abstract:

Mathematical models of infectious diseases exhibit robust dynamics such as, stable endemic or a disease-free equilibrium, or convergence of the solutions to periodic epidemic waves. The present works shows that the accuracy such dynamics can be significantly improved by incorporating both local and global dynamics of the infection in disease models. To demonstrate improved accuracies, we extended a standard Susceptible-Infected-Recovered (SIR) model by incorporating global dynamics of the COVID-19 pandemic. The extended SIR model assumes three possibilities for the susceptible individuals traveling outside of their community: They can return to the community without any exposure to the infection, they can be exposed and develop symptoms after returning to the community, or they can be tested positive during the trip and remain quarantined until fully recovered. To examine the predictive accuracies of the extended SIR model, we studied the prevalence of the COVID-19 infection in Kansas City, Missouri influenced by the COVID-19 global pandemic. Using a two-step model fitting algorithm, the extended SIR model was parametrized using the Kansas City, Missouri COVID-19 data during March to October 2020. The extended SIR model significantly outperformed the standard SIR model and revealed oscillatory behaviors with an increasing trend of infected individuals. In conclusion, the analytics and predictive accuracies of disease models can be significantly improved by incorporating the global dynamics of the infection in the models.

# 1- Introduction:

Mathematical modeling of infectious diseases has increasingly become an essential tool for prevention, prediction, and control of infectious diseases [1-4]. Since 1760, when Daniel Bernoulli developed the first disease model of smallpox, numerous mathematical models have been utilized to study disease transmission dynamics, and to predict, assess, and control infectious diseases [5-8].The substance of mathematical modeling lies in formulating a set of mathematical equations that mimic reality[9]. Mathematical models have been evolved from small sets of ordinary differential equations to sophisticated compartmental models with several equations (see [10-12] for a review).

One of the simplest, yet powerful, disease models is the standard Susceptible-Infected-Recovered (SIR) model, which was first introduced by Kermack and McKendrick in a series of three papers[13-15]. In a standard SIR model, the host population is divided into susceptible, infected and recovered individuals, denoted by S(t), I(t) and R(t), respectively. These quantities track the numbers of individuals in each compartment over different time periods [16-17]. The standard SIR model without birth and death is represented by the set of ordinary differential equations [18]:

$$\frac{dS}{dt}(t) = -\beta S(t)I(t)$$
$$\frac{dI}{dt}(t) = \beta S(t)I(t) - \gamma I(t) \quad (1)$$
$$\frac{dR}{dt}(t) = \gamma I(t)$$

Where $\beta$ is the average number of susceptible individuals infected by one infectious individual per contact per unit of time (the transmission rate), and $\gamma$ is the average number of infected individuals recovered per unit of time (recovery rate).

For decades, the standard SIR model has been extended to various forms by adding different compartments to suit the biological, spatio-temporal and social aspects of the disease dynamics or to study the impact of intervention strategies on the disease transmission dynamics in different communities[19-20]. For instance, it has been extended to SIR models with diffusion [21], contaminated environment [22-23], delay terms [24], several strains of infection [25], and multiple routes of infection [26].

The abovementioned extended SIR models contribute to the existing literatures. However, they largely ignore the effects global dynamics of infection on local communities. The presence of a global pandemic or a widespread infection can largely influence the dynamics of infection in a local community. It is therefore essential to include the global dynamics of infection in a disease model. There have been attempts to include the global dynamics in different SIR models [24,27]. Nevertheless, such extended SIR models have several unknown parameters and poorly fit to data of host population. Due to lack flexibility and poor fitness to data, there is a need for develop SIR models that are more practical. The present work aims to address this issue. We extend the SIR model to a new model that includes the global impacts of the infection and is also capable of fitting well to infectious disease data. To incorporate the global effect and test the predictive accuracies of the extended model we focus on the COVID-19 data in local communities of Kansas City, Missouri.

COVID-19 is the infectious disease caused by the severe acute respiratory syndrome novel coronavirus (SARS-CoV-2). Because the transmissibility of this virus is relatively high and the outbreaks remained undetected for several days, COVID-19 turned into a global pandemic. Almost all countries of the world have been exposed to this virus. Since January 2020, more 88 million individuals have become infected with the COVID-19. The infection has resulted more than 2 millions death as of February, 2021[28]. Just in the US , the COVID-19 cases are over 27 millions and more than 479,000 deaths as of February, 2021[29]. In order to reduce the spread of COVID-19, businesses, communities and governments have implemented different control measures such as mandatory lockdowns, social distancing, avoiding crowded events, and the use face masks in public [30]. Nevertheless, control of COVID-19 remains a major issue in several parts of the world [31].

COVID-19 is mainly transmitted from human-to-human via direct contact with contaminated surfaces and through the inhalation of respiratory droplets from infected individuals [32]. About 97% of the infected individuals will recover after period ranging between one to four weeks. Therefore, the use of SIR modeling seems to be an appropriate approach. Several researchers used mathematical modeling to analyze, and predict the transmission dynamics of COVID-19 pandemic [33-35]. Dynamics of COVID-19 epidemic has been simulated using different versions of SIR or SIER (susceptible, exposed, infected and recovered) models [33,34]. The main modification include adding asymptomatic and symptomatic infection compartments [35], hospitalization compartment [36], and quarantined and isolated compartments [33]. These models are presumably able to predict and simulate the number of infected cases by taking into consideration the asymptomatic and symptomatic cases, deaths, needs of beds in hospitals, and effect of control measures and the interventions to decrease the number of cases.

Although the abovementioned SIR models have been proven useful to study the dynamics of COVID-19, there are limitations due to the absence of global components. Moreover, regardless of the parameter values, most numerical simulations of SIR models are limited to three distinct dynamics. The first of these dynamics is the solution curve of infected individuals may exhibit an epidemic wave before converging to a disease-free equilibrium [37], secondly, the solution curve of infected converge to an endemic equilibrium [38], the third of these dynamics is the solution curve of infected converge to periodic epidemic waves [39]. For the standard SIR model (1), the dynamics are even more limited. Namely, the solution curves always represent the same qualitative dynamics: an epidemic wave of the infectious population, an inverted S shape for susceptible population, and S shape for the recovered population. Regardless of the set of parameter values and initial conditions, such qualitative behaviors will always remain the same (see panels Fig 1A- Fig 1C). A quick review of the number of individuals infected with COVID-19, at the country [28], State [40], or community level [41], shows that the dynamics of COVID-19 is more complicated than a single epidemic wave.

A missing, yet crucial component of the SIR models is the global effects of COVID-19 on the local communities. Most communities are well-connected and the assumption that the disease exists only within the community is invalid. There are three possibilities to consider in the SIR modeling of COVID-19: Susceptible individuals from a local community can travel in and out of their community without any exposure to COVID-19, they can be exposed to COVID-19 while traveling and develop symptoms after they return to their community, or they can be diagnosed with COVID-19 during their traveling and return to their community after recovery. In the next section we include all these three possibilities in the extended SIR model.

The rest of the paper is organized as follows. In section. 2, we introduce the extended SIR model and its formulation. We will also explain the methodology that was used to fit the extended SIR model to COVID-19 data. In section. 3, we present the results of our analysis and the comparison between the standard SIR and the extended SIR model. In section.4, we provide a discussion of the main results and additional factors to consider in the modeling process.

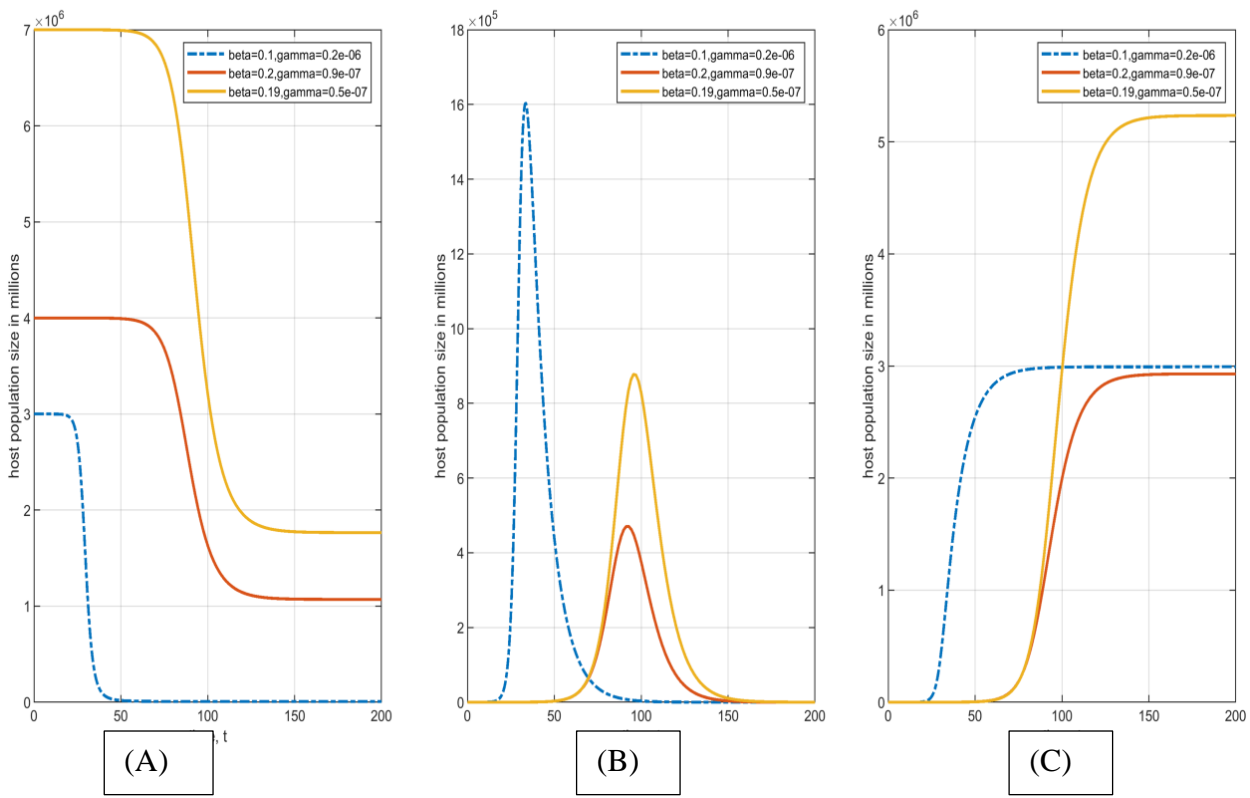

**Fig 1**. **Qualitative behavior of the standard SIR model remains the same regardless of the parameter values ($\beta$ and $\gamma$)**

(A) an inverted S-shape occurs for the susceptible population.

(B) a bell-shaped epidemic wave of infected population.

(C) a S-shaped curve of a recovered population

# Materials and Methods

## 2.1 Data

The COVID-19 data used in this study were obtained from the Kansas City Health Department [42]. The data were dated from March 14, 2020 to October 13, 2020 (a total of 212 days). Specifically, the data variables consisted of date, total number of cases, new cases, total deaths, new deaths and total number of individuals tested for COVID-19. We used abovementioned data to extract the daily number of recovered and susceptible and infected individuals (see S1 file for the algorithms used for generating the data).

Table (1) provides the basic descriptive statistics of Kansas City, Missouri daily COVID-19 data. Observe that the statistics of susceptible and recovered data are comparable. However, the statistics of infected individuals are at a much lower scale.

To estimate the number of susceptible individuals, we assumed an average incubation period of 5 days for COVID-19 [43]. We also considered one day for obtaining the COVID-19 test results. Hence, all of those who were tested positive were susceptible from the beginning until 6 days prior to obtaining the test results. Also, we added the individual who take the test, but their results were negative. These individuals had presumably high risk of getting infected and therefore susceptible. The number of infected individuals were calculated by considering an average infection period 14 days [44]. Hence, we cumulatively added of new cases for 14 days until they recovered (See the supplementary document for the algorithms used to calculate the number of infected and recovered individuals for each of 212 days).

**Table 1. Descriptive Statistics of Kansas City, Missouri daily COVID-19 data from March 14, 2020 to October 13, 2020**

|  | Susceptible | Infected | Recovered |
|---|---|---|---|
| Minimum | 91 | 1 | 0.90 |
| Maximum | 12,615 | 175 | 11,886.6 |
| Mean | 4,176.7 | 53.24 | 3,544.8 |
| Median | 2,325 | 36.5 | 1,677.5 |
| Standard Deviation | 3,670.7 | 49.16 | 3,832.9 |
| Range | 12,524 | 174 | 11,885.7 |

## 2.2 Model Formulation

We divided our population of N individuals living in a local community into sup-populations (i.e., compartments) of susceptible compartment S(t), infected compartment I(t), and recovered compartment R(t). As shown in Fig 2, the extended SIR model of COVID-19 transmission assumes three possibilities for susceptible individuals traveling outside of the community: They can return to the community without any exposure (the net rate is f(t)= $f_2$(t)-$f_1$(1)), they can be exposed COVID-19 and develop symptoms after returning to the community (the inflow rate of g(t)), or they can be tested positive during their trip and remain quarantined until fully recovered and thereafter return to the community (the inflow rate of h(t)). Then the extended SIR model is formulated by the following system of deterministic non-linear differential equations while Fig 2 gives the flow diagram of the model.

$$\frac{dS}{dt}(t) = -\beta S(t)I(t) + f(t)$$

$$\frac{dI}{dt}(t) = \beta S(t)I(t) - \gamma I(t) + g(t)$$

$$\frac{dR}{dt}(t) = \gamma I(t) + h(t),$$

(2)

where $\beta$ and $\gamma$ are the same parameters as in system (1). Functions $f(t), g(t)$ and $h(t)$ are differentiable and bounded and take into account the global effects of the infection. To avoid overfitting, our goal is to estimate $f, g$ and $h$ using least complicated forms.

Although adding exposed population can provide interesting dynamics, we decided to exclude the exposed compartment from our modeling. This is due to lack of data associated with exposed population. Namely, there is no known method of accurately identify the time series of exposed population in a community. In addition, the more compartments are added the harder it becomes to accurately estimate the parameter values. In some cases, the confidence intervals of estimated parameter values become extremely large due to high number of parameters and insufficient amount of data. With this rationale in mind, we therefore employ a two-step method to estimate the parameters of the model.

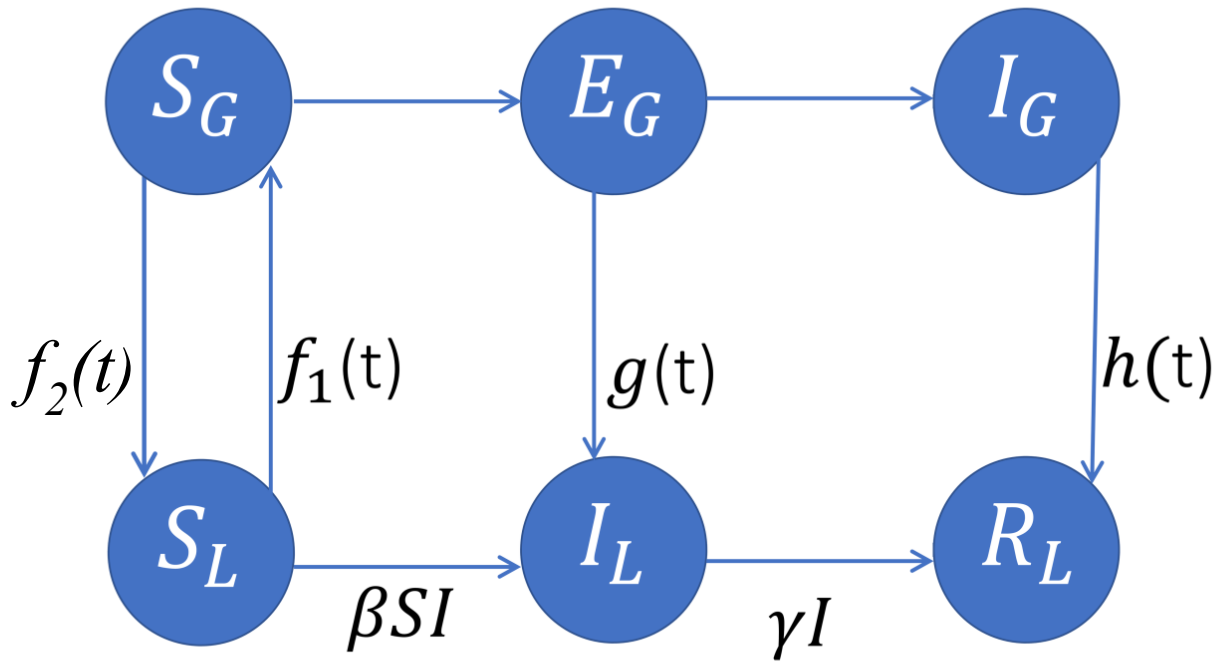

**Fig 2**. **A schematic representation of the extended SIR model coupled with a global SIER model.**

Individuals can get infected both within and outside of the community. A standard SIR model has been considered for progression of infection within the community. The extended model assumes three possibilities for susceptible individuals traveling outside of the community: They can return to the community without any exposure (the net rate is f(t)= $f_2(t)$-$f_1(t)$) ,they can be exposed to the infection and develop symptoms after returning to the community (rate of g(t)) or they can be tested positive during their trip and remain quarantined until fully recovered and thereafter return to the community (rate of h(t))

## 2.3 Model fitting

The single-step numerical methods such as linearization and discretization [45] to estimate the parameter values of model (2) fail to converge, due to high degrees of freedom and unknown intervals of parameter estimations. We therefore propose a two-step process for parameters estimation of model (2). We first estimate the parameter values of the standard SIR model (1) and then we determine functions f(t), g(t) and h(t) using the residual data of S(t), I(t) and R(t) subpopulations. As mentioned before, we used the COVID-19 data of susceptible, infected and recovered individuals in Kansas City, MO. for an epidemic period starting from March 12, 2020, to October 13, 2020.

In the first step, we numerically solved the system (2) using the MATLAB ode45 solver which is based on the fourth order Runge-Kutta method. The stability of the method is well established in [46]. In the same step, the model fitting was performed using *fmincon* function in the Optimization Toolbox of MATLAB. The *fmincon* function helps us to minimizes the sum of the squared error leading to the estimated parameter values of the SIR model under some constrains in order to get the best estimation for these parameters [47,49]. We estimated the SIR parameter values by considering the following factors. Several studies indicated that the COVID-19 transmission rate of infection was 0.5 [50,51]. Hence, we set $\beta = 0.5$. Also, some studies assumed the average recovery period (i.e $1/\gamma$) is about 7 days [50,51], which results in the initial value of $\gamma=0.13$. Also, to be consistent with the data, we set our initial conditions to S(0) = 12615, I(0) = 1, and R(0) = 300. The initial and estimated model parameters are provided in Table (2).

In the second step, we fitted the global effect of infection on the community by estimating functions f(t), g(t) and h(t) in model (2).

While several different functions can be considered for global effect fitting, to avoid any overfitting, we assumed:

$f(t) = \lambda.$

$g(t) = a_1 b_1 \cos(b_1 T + c_1) + a_2 b_2 \cos(b_2 T + c_2).$

$h(t) = 2P_1 t + P_2,$

where $\lambda$ is the net constant recruitment flow rate of susceptible individuals.

**Table 2. Initial and estimated parameters in the model (2)**

| Parameter | Description | Initial step 1 value | Initial step 2 value | Estimated values for step 1 | Estimated value for step 2 | Reference* |
|---|---|---|---|---|---|---|
| $\lambda$ | Constant recruitment rate | - | 10 | - | -2.6630 | Assumed |
| $\beta$ | transmission rate | 0.5 | 0.5 | 0.0816 | 0.0043 | [50,51] |
| $\gamma$ | Recovery rate. | 0.13 | 0.13 | 1.3e-11 | 2.8e-11 | [50,51] |

* The initial values were obtained from the indicated reference.

## 2- Results

Using the COVID-19 data, we estimated the functions corresponding to the global effects $f(t), g(t)$ and $h(t)$ in model (2) using the abovementioned two steps. The estimated net rate for the susceptible individual who can return to the community without any exposure is given by $f(t) = \lambda = $ -2.6630. The estimated net rate of the individual who exposed to the infection and develop symptoms after returning to the community is given by $g(t) = a_1 b_1 \cos(b_1 T + c_1) + a_2 b_2 \cos(b_2 T + c_2)$ where $a_1$=97.41 (95% CI: 83.4-111.4, $b_1$= 0.008485 (95% CI: 0.0065-0.015), $c_1$=2.759 (95% CI: 2.64-2.88), $a_2$= 24.12 (95% CI: 19.3-28.94), $b_2$= 0.04181 (95% CI: 0.037-0.047) and $c_2$= -2.359 (95% CI: -2.92,-1.80), and the estimated net rate of the individual who

tested positive during their trip and remain quarantined until fully recovered and thereafter return to the community is given by h(t)= $2P_1T+P_2$ where $P_1$= 0.3012 (95% CI: 0.291-0.312), and $P_2$= -22.02 (95% CI: -24.65,-19.39). All parameter estimations are summarized in Table 3.

**Table 3. Parameters of global effects functions in Infected and recovered models**

| Parameters | Estimation | 95%Confidence Interval (CI) |
|---|---|---|
| $a_1$ | 97.41 | (83.4, 111.4) |
| $b_1$ | 0.008485 | (0.0065, 0.015) |
| $c_1$ | 2.759 | (2.64, 2.88) |
| $a_2$ | 24.12 | (19.3, 28.94) |
| $b_2$ | 0.04181 | (0.037, 0.047) |
| $c_2$ | -2.359 | (-2.92, -1.80) |
| $P_1$ | 0.3012 | (0.291, 0.312) |
| $P_2$ | -22.02 | (-24.65, -19.39) |

Fig 3A and Fig 3B shows that the extended SIR model fits well to the data of susceptible, infection and recovered subpopulations. For the extended SIR model of the susceptible solution curve has, the $R^2 = 0.9905$ while in standard model $R^2 = 0.1551$. Similarly the extended SIR model outperformed the standard SIR model in the subpopulations of recovered individuals ($R^2 = 0.9912$ versus $R^2 = 0.47$), and the subpopulation of infected individuals ($R^2 = 0.7083$ versus $R^2 = -258.65$). Note that the negative $R^2$ value is because the classical SIR model does not follow the trend of the data (see Fig 4A and Fig 4B in the discussion section ).This is due to the sum square of the residual (SSE) being greater than the sum square of the total (SST). It is important to note that regardless of numerical method of model fitting (e.g., fminsearch alghorim) the standard SIR model has an extremely poor fitting with respect to the infected individuals. As summarized in Table 4, the extended SIR model has significantly higher predictive accuracies than the standard SIR model in the terms of $R^2$, Akaike's Information Criterion (AIC) and corrected AIC. Therefore, there are significant improvements to goodness of fit and predictive accuracies by including the global effects.

In Fig 3A observed that there are two periodic waves of total infected individuals. The first wave of infected cases was on March 25 which is the day 13 after the first case of COVID-19 recorded. Also, It was a day after staying home order that issued by Kansas City government in order to decrease the number of COVID-19 cases [52]. The infected cases were reached the second wave on August 10. This wave may relate the summertime where all the schools were closed, and the families were out of the town and return home. That shows how important consider the global effects in the SIR model because several individuals can get infected as they are traveling outside of their local communities due to their work or study and thereafter return to the community.

**Table 4. Comparisons of model fitness for the standard and extended SIR models**

| Fitness | Extended | Standard |
|---|---|---|
| Corrected AIC | 3.2799e+03 | 4.3130e+03 |
| AIC | 3.0351e+03 | 4.0689e+03 |
| SSR ($R^2$) for S | 3.22e+09 (0.9905) | 5.0377e+08 (0.1551) |
| SSR ($R^2$) for I | 4.1249e+05 (0.7083) | -1.5064e+08 (-258.65)* |
| SSR ($R^2$) for R | 3.51e+09 (0.9912) | 1.6640e+09 (0.4700) |

\* The standard SIR model has an extremely poor fitting with respect to the infected individuals.

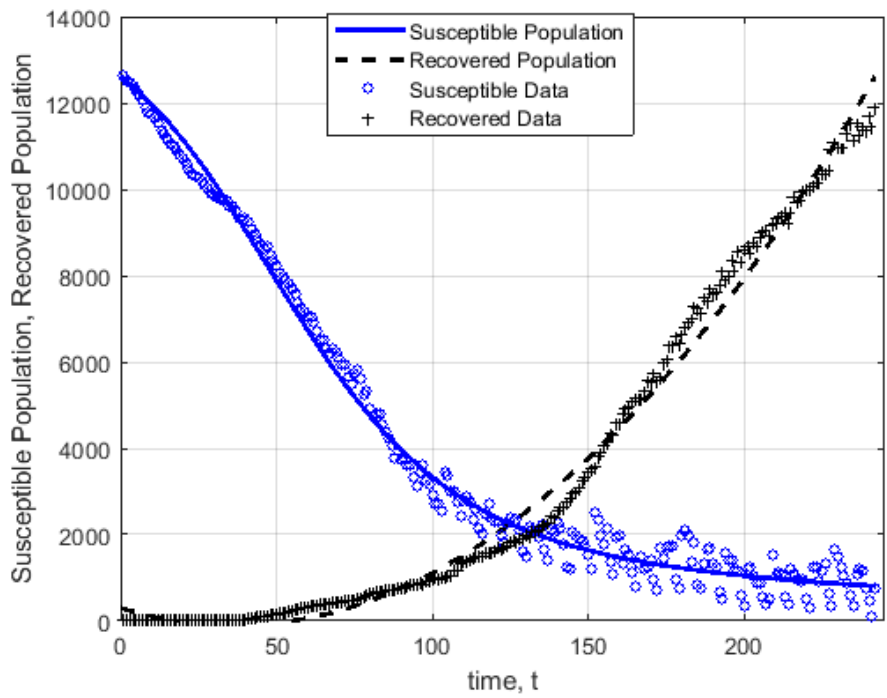

**(A)**

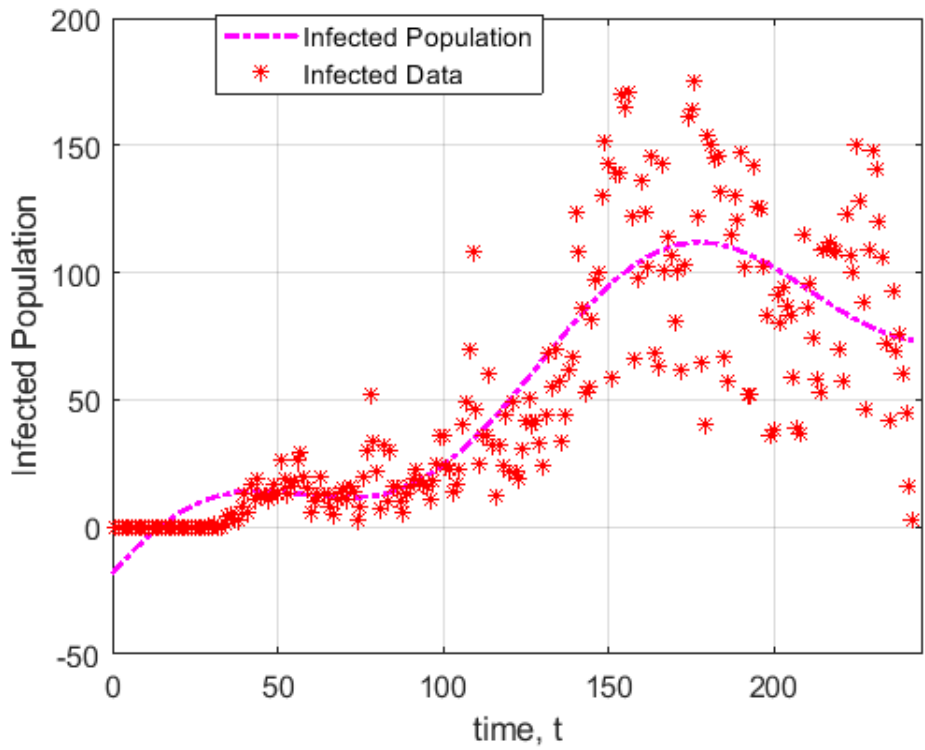

**(B)**

**Fig 3**. **The extended SIR model (2) has a relatively strong goodness of fit.**

(A) The extended SIR model fitted to the COVID-19 data of susceptible and recovered subpopulation in Kansas City, Missouri.

(B) The extended SIR model fitted to the COVID-19 data of infected subpopulation in Kansas City, Missouri.

## 4- Discussion:

The present work highlights the importance of including global dynamics of infection in disease models to achieve higher predictive accuracies. We introduced a two-step algorithm for accurate estimation of infection parameters by considering both global and local effects of the infection spread in a disease model. The first step leads to estimation of local parameters (i.e., the transmission and recovery rates, $\beta$ and $\gamma$, respectively) whereas the second step incorporates the global effects of the infection (i.e., estimation of functions $f(t), g(t)$ and $h(t)$). To test the methodology, we applied the two-step model fitting algorithm to the extended SIR model (2) using the Kansas City COVID-19 data from March to October 2020. As shown in Fig 3A and Fig 3B the two-step method resulted in solution curves that fit well to the COVID-19 data. The goodness of fit becomes more apparent when it is compared to that of the standard SIR model (1). Although the standard SIR has been proven useful to study local dynamics of various infections, it fails to capture the global effects of a widespread disease (see Table 4 to observe its poor statistical fitness). As shown in Fig 4A and Fig 4B, the solution curves of the standard SIR model poorly fit to the COVID-19. In addition to higher predictive accuracies of the extended SIR model (2), the solution curves revealed oscillatory behaviors with an increasing trend of infected individuals. This contrasts with the standard SIR model, where regardless of chosen parameter values, the solution curves always exhibit the same qualitative behaviors (see Fig1A-Fig 1C).

Particularly, in Fig 3B for the extended SIR model, we notice two epidemic waves of infected individuals in Kansas City, Missouri. The first wave of infected cases was approximately on March 25 which is the day 13 after the first case of COVID-19 recorded in Kansas City. Also, it was a few days after stay-at-home order issued by the Kansas City government to decrease the number of COVID-19 cases [52]. The infected cases were reached to the second wave approximately on August 10. This wave may relate the summertime where all the schools were closed, and families traveling outside of Kansas City returned home. Hence, by including the global infection effects in the disease models, we can identify underlying mechanisms governing the dynamics of infectious diseases. Namely, several individuals traveling outside of their local communities may have returned as infected or exposed individuals.

The inability of the standard SIR model to fit there COVID-19 data has been identified by other researchers [53]. Nonetheless, the presence of a breakpoint due to strong policy interventions, mentioned in [53], does not necessarily reduce the prevalence of infection in a community dealing with a pandemic. For instance, in Kansas City, Missouri, the stay-at-home order did not result in a substantial reduction in the number of COVID infected individuals. The proposed methodology in this study can be applied to a variety of disease models.

In conclusion, Including the global dynamics of the infection and applying the two-step model fitting algorithm can enable us to extract vital information (e.g., presence of epidemic waves) from the data.

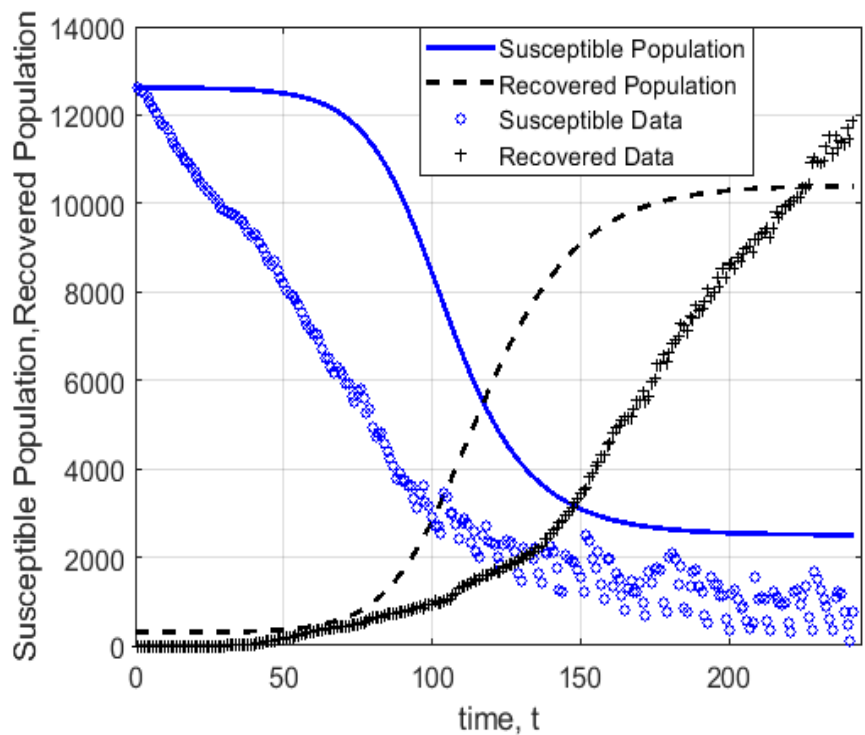

**(A)**

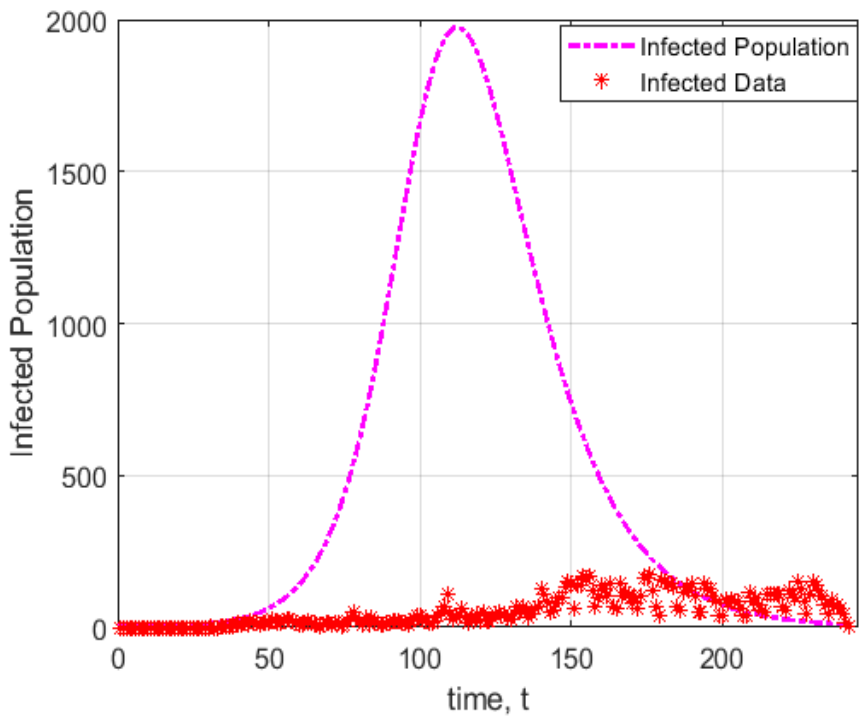

**(B)**

**Fig 4**. **The standard SIR model fitted to fitted COVID-19 data.**

(A) The standard SIR model poorly fits to the susceptible and recovered data.

(B) The standard SIR model has poor fitness to data of infected population.

# References:

1. Nwankwo A, Okuonghae D. A Mathematical Model for the Population Dynamics of Malaria with a Temperature Dependent Control. Differ Equations Dyn Syst [Internet]. 2019; Available from: https://doi.org/10.1007/s12591-019-00466-y

2. Keeling MJ, Danon L. Mathematical modelling of infectious diseases. Br Med Bull [Internet]. 2009 Dec 1;92(1):33–42. Available from: https://doi.org/10.1093/bmb/ldp038

3. Huppert A, Katriel G. Mathematical modelling and prediction in infectious disease epidemiology. Clin Microbiol Infect [Internet]. 2013;19(11):999–1005. Available from: http://www.sciencedirect.com/science/article/pii/S1198743X14630019

4. mathematical-modeling @ www.cdc.gov [Internet]. Available from: https://www.cdc.gov/coronavirus/2019-ncov/covid-data/mathematical-modeling.html

5. Roberts MG, Heesterbeek J a P. Mathematical models in epidemiology. Math Model. 2003;49(0):6221.

6. Siettos CI, Russo L. Mathematical modeling of infectious disease dynamics. Virulence [Internet]. 2013 May 15;4(4):295–306. Available from: https://doi.org/10.4161/viru.24041

7. Kretzschmar M, Wallinga J. Mathematical Models in Infectious Disease Epidemiology. Krämer A, Kretzschmar M, Krickeberg K, editors. Mod Infect Dis Epidemiol Concepts, Methods, Math Model Public Heal [Internet]. 2009 Jul 28;209–21. Available from: https://www.ncbi.nlm.nih.gov/pmc/articles/PMC7178885/

8. Brauer F. Mathematical epidemiology: Past, present, and future. Infect Dis Model [Internet]. 2017;2(2):113–27. Available from: http://www.sciencedirect.com/science/article/pii/S2468042716300367

9. Panovska-Griffiths J. Can mathematical modelling solve the current Covid-19 crisis?

BMC Public Health [Internet]. 2020;20(1):551. Available from: https://doi.org/10.1186/s12889-020-08671-z

10. Roddam AW. Mathematical Epidemiology of Infectious Diseases: Model Building, Analysis and Interpretation: O Diekmann and JAP Heesterbeek, 2000, Chichester: John Wiley pp. 303, £39.95. ISBN 0-471-49241-8. Int J Epidemiol [Internet]. 2001 Feb 1;30(1):186. Available from: https://doi.org/10.1093/ije/30.1.186

11. Edoh K, MacCarthy E. Network and equation-based models in epidemiology. Int J Biomath. 2018 Mar 19;11.

12. Hejblum G, Setbon M, Temime L, Lesieur S, Valleron A-J. Modelers' Perception of Mathematical Modeling in Epidemiology: A Web-Based Survey. PLoS One [Internet]. 2011 Jan 31;6(1):e16531. Available from: https://doi.org/10.1371/journal.pone.0016531

13. Kermack WO, McKendrick AG. Contributions to the mathematical theory of epidemics—I. Bull Math Biol [Internet]. 1991;53(1):33–55. Available from: https://doi.org/10.1007/BF02464423

14. Kermack WO, McKendrick AG. Contributions to the mathematical theory of epidemics—II. The problem of endemicity. Bull Math Biol [Internet]. 1991;53(1):57–87. Available from: https://doi.org/10.1007/BF02464424

15. Kermack WO, McKendrick AG. Contributions to the mathematical theory of epidemics—III. Further studies of the problem of endemicity. Bull Math Biol [Internet]. 1991;53(1):89–118. Available from: https://doi.org/10.1007/BF02464425

16. Chen D. Modeling the Spread of Infectious Diseases: A Review. In 2014. p. 19–42.

17. Shrestha S, Lloyd-Smith J. Introduction to mathematical modeling of infectious diseases. 2010;0000:1–46.

18. Harko T, Lobo FSN, Mak MK. Exact analytical solutions of the Susceptible-Infected-Recovered (SIR) epidemic model and of the SIR model with equal death and birth rates. Appl Math Comput [Internet]. 2014;236:184–94. Available from: http://www.sciencedirect.com/science/article/pii/S009630031400383X

19. Kim S, Lee J, Jung E. Mathematical model of transmission dynamics and optimal control strategies for 2009 A/H1N1 influenza in the Republic of Korea. J Theor Biol [Internet]. 2017;412:74–85. Available from: http://www.sciencedirect.com/science/article/pii/S0022519316303228

20. Del Valle S, Hethcote H, Hyman JM, Castillo-Chavez C. Effects of behavioral changes in a smallpox attack model. Math Biosci [Internet]. 2005;195(2):228–51. Available from: http://www.sciencedirect.com/science/article/pii/S0025556405000593

21. Gai C, Iron D, Kolokolnikov T. Localized outbreaks in an S-I-R model with diffusion. J Math Biol [Internet]. 2020;80(5):1389–411. Available from: https://doi.org/10.1007/s00285-020-01466-1

22. GAUTAM R, LAHODNY G, BANI-YAGHOUB M, MORLEY PS, IVANEK R. Understanding the role of cleaning in the control of Salmonella Typhimurium in grower-finisher pigs: a modelling approach. Epidemiol Infect [Internet]. 2013/08/07. 2014;142(5):1034–49. Available from: https://www.cambridge.org/core/article/understanding-the-role-of-cleaning-in-the-control-of-salmonella-typhimurium-in-growerfinisher-pigs-a-modelling-approach/0583A9978A127B4F47482EE6A8B048DF

23. Gautam R, Bani-Yaghoub M, Neill WH, Döpfer D, Kaspar C, Ivanek R. Modeling the effect of seasonal variation in ambient temperature on the transmission dynamics of a pathogen with a free-living stage: example of Escherichia coli O157:H7 in a dairy herd. Prev Vet Med [Internet]. 2011;102(1):10–21. Available from: http://europepmc.org/abstract/MED/21764472

24. Zhang S-P, Yang Y-R, Zhou Y-H. Traveling waves in a delayed SIR model with nonlocal dispersal and nonlinear incidence. J Math Phys [Internet]. 2018 Jan 1;59:11513. Available from: https://ui.adsabs.harvard.edu/abs/2018JMP....59a1513Z

25. Bani-Yaghoub M, Gautam R, Shuai Z, van den Driessche P, Ivanek R. Reproduction numbers for infections with free-living pathogens growing in the environment. J Biol Dyn [Internet]. 2012 Mar 1;6(2):923–40. Available from: https://doi.org/10.1080/17513758.2012.693206

26. BANI-YAGHOUB M, GAUTAM R, DÖPFER D, KASPAR CW, IVANEK R. Effectiveness of environmental decontamination as an infection control measure. Epidemiol Infect [Internet]. 2011/05/18. 2012;140(3):542–53. Available from: https://www.cambridge.org/core/article/effectiveness-of-environmental-decontamination-as-an-infection-control-measure/FA961694D9AAE47733792629750871 91

27. Li W-T, Lin G, Ma C, Yang F-Y. Traveling wave solutions of a nonlocal delayed SIR model without outbreak threshold. Discret Contin Dyn Syst - B [Internet]. 19(2):467–84. Available from: http://aimsciences.org//article/id/d99e6db9-456f-4a01-add9-e0b99b823bb4

28. Coronavirus @ Ourworldindata.Org [Internet]. Available from: https://ourworldindata.org/coronavirus

29. c33a67efa088af42cf95f022a7fdde636782c817 @ covid.cdc.gov [Internet]. Available from: https://covid.cdc.gov/covid-data-tracker/#cases_casesper100klast7days

30. Chu DK, Akl EA, Duda S, Solo K, Yaacoub S, Schünemann HJ, Chu DK, Akl EA, El-harakeh A, Bognanni A, Lotfi T, Loeb M, Hajizadeh A, Bak A, Izcovich A, Cuello-Garcia CA, Chen C, Harris DJ, Borowiack E, Chamseddine F, Schünemann F, Morgano GP, Muti Schünemann GEU, Chen G, Zhao H, Neumann I, Chan J, Khabsa J, Hneiny L, Harrison L, Smith M, Rizk N, Giorgi Rossi P, AbiHanna P, El-khoury R, Stalteri R, Baldeh T, Piggott T, Zhang Y, Saad Z, Khamis A, Reinap M, Duda S, Solo K, Yaacoub S, Schünemann HJ. Physical distancing, face masks, and eye protection to prevent person-to-person transmission of SARS-CoV-2 and COVID-19: a systematic review and meta-analysis. Lancet [Internet]. 2020;395(10242):1973–87. Available from: http://www.sciencedirect.com/science/article/pii/S0140673620311429

31. 6aec8563b3a15c84c7093cbaca99ef5cdad2ea3b @ www.washingtonpost.com [Internet]. Available from: https://www.washingtonpost.com/business/2020/03/12/markets-live-updates-coronavirus-economy/

32. Li Q, Guan X, Wu P, Wang X, Zhou L, Tong Y, Ren R, Leung KSM, Lau EHY, Wong JY, Xing X, Xiang N, Wu Y, Li C, Chen Q, Li D, Liu T, Zhao J, Liu M, Tu W, Chen C, Jin L, Yang R, Wang Q, Zhou S, Wang R, Liu H, Luo Y, Liu Y, Shao G, Li H, Tao Z, Yang Y, Deng Z, Liu B, Ma Z, Zhang Y, Shi G, Lam TTY, Wu JT, Gao GF, Cowling BJ, Yang B, Leung GM, Feng Z. Early Transmission Dynamics in Wuhan, China, of Novel Coronavirus–Infected Pneumonia. N Engl J Med. 2020;382(13):1199–207.

33. Ivorra B, Ferrández MR, Vela-Pérez M, Ramos AM. Mathematical modeling of the spread of the coronavirus disease 2019 (COVID-19) taking into account the undetected infections. The case of China. Commun Nonlinear Sci Numer Simul [Internet]. 2020;88:105303. Available from:

http://www.sciencedirect.com/science/article/pii/S1007570420301350

34. Kim S, Seo Y Bin, Jung E. Prediction of COVID-19 transmission dynamics using a mathematical model considering behavior changes in Korea . Epidemiol Heal [Internet]. 2020 Apr 13;42(0):e2020026-0. Available from: https://doi.org/10.4178/epih.e2020026

35. Tomochi M, Kono M. A mathematical model for COVID-19 pandemic—SIIR model: Effects of asymptomatic individuals. J Gen Fam Med [Internet]. 2020 Nov 1;n/a(n/a). Available from: https://doi.org/10.1002/jgf2.382

36. Alshammari FS. A Mathematical Model to Investigate the Transmission of COVID-19 in the Kingdom of Saudi Arabia. Yetilmezsoy K, editor. Comput Math Methods Med [Internet]. 2020;2020:9136157. Available from: https://doi.org/10.1155/2020/9136157

37. Köhler-Rieper F, Röhl CHF, De Micheli E. A novel deterministic forecast model for the Covid-19 epidemic based on a single ordinary integro-differential equation. Eur Phys J Plus [Internet]. 2020;135(7):599. Available from: https://doi.org/10.1140/epjp/s13360-020-00608-0

38. Sirijampa A, Chinviriyasit S, Chinviriyasit W. Hopf bifurcation analysis of a delayed SEIR epidemic model with infectious force in latent and infected period. Adv Differ Equations [Internet]. 2018;2018(1):348. Available from: https://doi.org/10.1186/s13662-018-1805-6

39. Oluyori DA. Backward and Hopf bifurcation analysis of an SEIRS COVID-19 epidemic model with saturated incidence and saturated treatment response. 2020;1–26. Available from: https://www.medrxiv.org/content/10.1101/2020.08.28.20183723v1.full.pdf

40. Wu SL, Mertens AN, Crider YS, Nguyen A, Pokpongkiat NN, Djajadi S, Seth A, Hsiang MS, Colford JM, Reingold A, Arnold BF, Hubbard A, Benjamin-Chung J. Substantial underestimation of SARS-CoV-2 infection in the United States. Nat Commun [Internet]. 2020;11(1):4507. Available from: https://doi.org/10.1038/s41467-020-18272-4

41. MARC. https://marc2.org/covidhub/ [Internet]. Available from: https://marc2.org/covidhub/

42. nfta-sjx6 @ data.kcmo.org [Internet]. Available from: https://data.kcmo.org/Health/COVID-19-Case-Death-Trends-by-Date/nfta-sjx6

43. Rothan HA, Byrareddy SN. The epidemiology and pathogenesis of coronavirus disease (COVID-19) outbreak. J Autoimmun [Internet]. 2020;109:102433. Available from: http://www.sciencedirect.com/science/article/pii/S0896841120300469

44. Qin J, You C, Lin Q, Hu T, Yu S, Zhou X-H. Estimation of incubation period distribution of COVID-19 using disease onset forward time: A novel cross-sectional and forward follow-up study. Sci Adv [Internet]. 2020 Aug 1;6(33):eabc1202. Available from: http://advances.sciencemag.org/content/6/33/eabc1202.abstract

45. Ramsay JO, Hooker G, Campbell D, Cao J. Parameter estimation for differential equations: a generalized smoothing approach. J R Stat Soc Ser B (Statistical Methodol [Internet]. 2007 Nov 1;69(5):741–96. Available from: https://doi.org/10.1111/j.1467-9868.2007.00610.x

46. May R, Noye J. The Numerical Solution of Ordinary Differential Equations: Initial Value Problems. In: Noye J, editor. Computational Techniques for Differentail Equations [Internet]. North-Holland; 1984. p. 1–94. (North-Holland Mathematics Studies; vol. 83). Available from: http://www.sciencedirect.com/science/article/pii/S0304020808712003

47. Byrd RH, Hribar ME, Nocedal J. An Interior Point Algorithm for Large-Scale Nonlinear Programming. SIAM J Optim [Internet]. 1999 Jan 1;9(4):877–900. Available from:

https://doi.org/10.1137/S1052623497325107

48. fmincon @ www.mathworks.com [Internet]. Available from: https://www.mathworks.com/help/optim/ug/fmincon.html

49. Waltz RA, Morales JL, Nocedal J, Orban D. An interior algorithm for nonlinear optimization that combines line search and trust region steps. Math Program [Internet]. 2006;107(3):391–408. Available from: https://doi.org/10.1007/s10107-004-0560-5

50. Cauchemez S, Fraser C, Van Kerkhove MD, Donnelly CA, Riley S, Rambaut A, Enouf V, van der Werf S, Ferguson NM. Middle East respiratory syndrome coronavirus: quantification of the extent of the epidemic, surveillance biases, and transmissibility. Lancet Infect Dis [Internet]. 2014;14(1):50–6. Available from: http://www.sciencedirect.com/science/article/pii/S1473309913703049

51. Chen T-M, Rui J, Wang Q-P, Zhao Z-Y, Cui J-A, Yin L. A mathematical model for simulating the phase-based transmissibility of a novel coronavirus. Infect Dis Poverty [Internet]. 2020;9(1):24. Available from: https://doi.org/10.1186/s40249-020-00640-3

52. https://www.kcmo.gov/ [Internet]. Available from: https://www.kcmo.gov/city-hall/departments/health/coronavirus-covid-19/stay-at-home-order-faq

53. Kim YJ, Seo MH, Yeom HE. Estimating a breakpoint in the pattern of spread of COVID-19 in South Korea. Int J Infect Dis. 2020 Aug 1;97:360–4.

# Supporting information

**S1 File. The algorithm for generating infected and recovered data.**